# DADL: A Declarative Description Language for Enterprise Tool Libraries in LLM Agent Systems

**Axel Dunkel**
Dunkel Cloud GmbH, Hofheim am Taunus, Germany
ad@dunkel.cloud
ORCID: 0009-0007-5885-354X

*April 2026*

## Abstract

The Model Context Protocol (MCP) has rapidly established itself as the standard interface between large language model (LLM) agents and external tools. Adopting MCP at organizational scale, however, exposes two structural problems. First, every API integration is shipped as a dedicated server process with its own deployment, dependency tree, and credential handling, and recent empirical work shows that the overwhelming majority of these servers are thin wrappers around REST APIs. Second, the per-tool registration model causes context window consumption to grow linearly with the number of available tools, which forces real deployments to expose only a small fraction of the APIs an organization actually uses. We present **DADL** (Dunkel API Description Language), a YAML format that describes a REST API's endpoints, authentication, pagination, response shaping, and access classification in a single declarative file. A DADL file is interpreted by an execution layer at runtime; no per-API server process is deployed and no integration code is generated, although the shared runtime is itself a server. Because all tools share that one runtime, credentials are managed centrally, authorization decisions are made centrally, and the available tools are exposed to the LLM through a fixed-size *Code Mode* interface that is independent of catalog size. The combination produces what we call an **Enterprise Tool Library**: a versioned, auditable collection of API integrations that any team in an organization can extend, share, and consume through a common authentication and authorization boundary. The DADL v0.1 specification is released under CC BY-SA 4.0, and a public registry currently contains 1,833 tool definitions across 20 services. On this catalog, Code Mode reduces the LLM context cost of *tool advertisement* — the schemas the model must see in order to know what is available — from approximately 142,000 tokens to approximately 1,000, a 142x reduction; the per-call cost of `search` and `execute` invocations is additional and depends on the task.

# 1. Introduction

The integration of LLM agents with external systems has moved from prototype to production. OpenAI's function-calling interface in 2023 [1] and Anthropic's Model Context Protocol (MCP) in late 2024 [2] standardized how a language model invokes external tools. MCP in particular has been adopted by major IDEs and AI assistants and is now a de facto interoperability layer for tool-augmented AI [3].

Two structural issues follow from MCP's current shape. The first is **server proliferation**. Each integration is packaged as a standalone server that implements an MCP transport, declares tool schemas, performs HTTP calls, handles authentication, and shapes responses. Mastouri et al. [4] analyzed 116 official MCP servers and found that 88.6 % are fully or partially backed by REST APIs and 92 % of their tools function as bare API wrappers. The same study reports that the median server exposes only 19 % of the underlying API operations. Hasan et al. [5] independently confirm pervasive maintainability and security weaknesses across thousands of community servers, and Radosevich & Halloran [6] demonstrate concrete exploitation paths in widely deployed MCP servers, including credential leakage and command injection. The ecosystem is therefore investing significant engineering effort to produce hand-written HTTP request templates that are simultaneously incomplete and insecure.

The second issue is **context window consumption**. MCP's current convention requires every tool's name, description, and parameter schema to be present in the LLM's context for every interaction. An API such as the GitHub REST API exposes more than 900 endpoints; even a 200-endpoint subset consumes tens of thousands of tokens before any user input. Recent work on dynamic tool memory management for multi-turn agent conversations [7] explicitly identifies this as a first-order constraint on agent capability. Consequently, organizations either run a few large MCP servers with severely truncated coverage or fragment their tooling across many sessions, neither of which matches how internal users actually work.

These two issues compound in enterprise deployments. A typical organization uses dozens of internal and SaaS APIs. Building one MCP server per API leads to dozens of independently deployed processes, each with its own credential storage and access control logic. Different teams independently rebuild the same wrappers for the same APIs, producing inconsistent interfaces and parallel attack surfaces. The result is the worst of both worlds: high operational cost and incomplete, fragmented coverage of the API landscape.

This paper presents an alternative model. Rather than maintain one MCP server process per API, organizations write a **DADL file** per API and run them all behind a single shared runtime: a YAML document declares the endpoints, the authentication scheme, pagination, response transformation, and access classification, and the runtime interprets that document at request time, performs the HTTP call, applies transformations, enforces authorization, and returns the result. DADL does not remove the need for a server process — the runtime is itself a server — but it removes the need for *one server per API* and the operational, credential, and authorization fragmentation that follows. From the LLM's perspective, the entire catalog is reachable through a single fixed-size *Code Mode* interface (Section 4). From the organization's perspective, all DADL files form a versioned, reviewable library that lives next to other infrastructure code.

The contributions of this paper are:

1. **DADL v0.1**, a declarative YAML specification for REST API integration that covers five authentication schemes, four pagination strategies, jq-based response transformation, an access classification system with four well-known levels, an embedded *hints* mechanism for institutional API knowledge, and a constrained composite-tool feature for cross-endpoint workflows (Section 3).
2. **Code Mode**, an execution model that exposes an arbitrary number of DADL-defined tools to the LLM

through two MCP-level operations (`search` and `execute`), holding tool-advertisement context cost approximately constant as the catalog grows (Section 4).

3. **The Enterprise Tool Library pattern** (Section 5): a deployment model in which DADL files are managed as versioned configuration, share a central credential store and authorization layer, encode institutional knowledge through hints and composite tools, and serve as living documentation for an organization's API surface. Within this pattern we name the **Library Composition Effect** (Section 5.5), the emergent capability of an LLM agent to construct cross-source queries against a uniformly-described tool catalog in response to questions never anticipated by the library's authors.
4. **A public registry** of 1,833 tools across 20 services, released under CC BY-SA 4.0, framed as a research artifact: a uniform machine-readable corpus of REST-API descriptions in a single declarative format, suitable for empirical studies of tool-use, context-cost measurement, and translation benchmarking (Section 6).

The DADL specification is released under Creative Commons Attribution-ShareAlike 4.0 (CC BY-SA 4.0). A public registry of community-contributed DADL files is available at [https://dadl.ai](https://dadl.ai). The reference execution layer is described only briefly, since its detailed architecture is the subject of separate work.

---

# 2. Background

## 2.1 The Model Context Protocol

MCP defines a bidirectional, schema-driven protocol between an LLM-hosting client and a tool server [2]. The client discovers a server's tools through a `list_tools` request that returns each tool's name, description, and JSON Schema for parameters. The client then invokes a tool with `call_tool`. Transports include STDIO and HTTP+SSE; the wire format is JSON-RPC 2.0.

The protocol is intentionally minimal: it standardizes the *interaction* between client and tool, not the *content* of the tool catalog. Authentication, authorization, response shaping, and pagination are entirely the responsibility of each server.

## 2.2 The Wrapper Problem

Mastouri et al. [4] empirically characterize the dominant server pattern: thin wrappers around REST APIs. Their AutoMCP pipeline demonstrates that MCP server code can be generated automatically from OpenAPI specifications, achieving a 76 % initial success rate, raised to 94.2 % after automated repair. AutoMCP confirms the diagnosis — most servers are mechanical translations of REST documentation — but its remedy is to *generate* code, leaving each generated server with the same deployment, dependency, and per-tool context cost as a hand-written one.

Li et al. [8] present ToolFactory, which generates tool definitions from unstructured documentation rather than OpenAPI. The output is again executable code, with the same operational properties. MCP Bridge [9] takes a complementary approach by aggregating multiple MCP servers behind a single HTTP proxy; it addresses deployment topology but not the existence of the underlying servers or their context cost.

## 2.3 Context Window Scalability

Because MCP advertises tools by listing each tool's full schema, the cost of having a tool *available* (regardless of whether it is called) is proportional to the number of tools. For modern models with 200K-token context windows, this becomes the binding constraint long before reasoning capacity does. A single GitHub MCP server covering ~200 endpoints can consume on the order of 50,000 tokens of context for tool schemas alone, leaving little room for conversation, code, or other tools. In practice this forces curators to expose narrow tool subsets, which contradicts the goal of giving an agent broad organizational coverage.

## 2.4 Related Description Formats

A declarative description of an API is not a new idea, and DADL must be located among several adjacent traditions.

**OpenAPI** (formerly Swagger) [10] is the dominant API description format for human and machine consumers. OpenAPI is, however, designed to *document* an API, not to specify how an LLM should interact with one: it has no concept of access classification, no notion of caller-side response transformation for context efficiency, no facility for embedding institutional knowledge ("call X before Y to obtain Z"), and no mechanism for executing the resulting description directly. DADL is intentionally complementary: a DADL file may import an OpenAPI spec for type definitions, but adds the LLM-oriented concerns OpenAPI omits.

**Postman Collections** capture executable HTTP requests with environment variables and pre/post scripts. They are excellent for human-driven exploration but were not designed as a machine-readable interface for autonomous agents; their imperative scripting model is not auditable in the sense of Property 1 (Section 3.11).

**AsyncAPI** describes event-driven APIs (WebSocket, MQTT, Kafka). Its scope is orthogonal to DADL, which addresses the request-response REST pattern. AsyncAPI-style streaming APIs are an explicit non-target for DADL v0.1 (Section 7.2).

**Terraform provider schemas** are perhaps the closest cousin: declarative, machine-readable descriptions of remote resources, with credential references resolved at runtime, executed by a shared runtime (Terraform), and assembled into reusable libraries (the provider registry). DADL borrows the same operational shape but targets the request-driven semantics of LLM tool calls rather than the desired-state semantics of infrastructure provisioning.

**Low-code connector frameworks** such as the Airbyte Low-Code CDK [11] (whose pagination vocabulary DADL adopts), Zapier-style integrations, and the connector definitions of integration platforms describe REST APIs declaratively for ETL or workflow purposes. They typically do not address access classification for autonomous agents and do not target context-efficient exposure to an LLM.

**API gateway configurations** (Kong, Apigee, Tyk, AWS API Gateway) declare routing, authentication, transformation, and rate-limiting in YAML or JSON, with a shared runtime executing the resulting policies. DADL shares this operational shape but moves up the stack: instead of declaring how to *expose* an API to clients, it declares how an LLM-side agent should *call* the upstream API, with vocabulary for context efficiency, access classification, and embedded domain knowledge that gateway formats lack.

DADL's distinct contribution is the combination: declarative like OpenAPI, runtime-interpreted like Terraform providers and gateway configs, with first-class affordances for the constraints of LLM tool use (context cost, access classification, hints, composite reuse).

---

# 3. The DADL Specification

## 3.1 Design Principles

DADL is guided by four principles.

**Declarative.** A DADL file specifies *what* endpoints exist, *how* to authenticate, and *how* responses should be shaped. There is no control flow in the base language. A reviewer reading the file can predict the complete set of HTTP endpoints any tool can address.

**Auditable HTTP surface.** Because the format contains no executable code in the common case, the set of HTTP endpoints reachable through a DADL file is determinable by static inspection. We state this property formally in Section 3.11 as *Declarative Closure*. This is a structural guarantee about the file's HTTP behavior, not a claim that any DADL file is comprehensible at a glance: a 2,700-line GitHub DADL with 200 tools, jq filters, and cross-file includes still requires careful review. What Declarative Closure provides is that the review is *bounded* — there is no hidden code path that can construct a request the file does not name. The single exception, composite tools (Section 3.7), is explicitly flagged with `contains_code: true` and is subject to additional review.

**Credential-free by construction.** DADL files contain logical credential references (e.g. `vault/stripe-secret-key`), never secret values. This makes definitions safe to commit, share, and review.

**LLM-generatable.** The format was deliberately chosen to be producible by an LLM from API documentation in a single interaction. This matters because it makes the cost of *adding* a new API to the library small enough that gaps in coverage are addressable by individual contributors.

## 3.2 File Structure

A DADL file uses the `.dadl` extension and is a YAML document with a top-level `backend` object and metadata fields:

```yaml
spec: "https://dadl.ai/spec/dadl-spec-v0.1.md"
credits: ["Jane Doe (@janedoe)"]
source_name: "Example REST API"
source_url: https://docs.example.com/api
date: "2026-03-26"

backend:
  name: my-api
  type: rest
  version: "1.0"
  base_url: https://api.example.com/v1
  description: "My REST API"

  auth:
    type: bearer
    credential: vault/my-api-token

  tools:
    list_items:
      method: GET
```

```
      path: /items
      access: read
      description: "List all items"
```

**Listing 1.** A minimal DADL file. The full vocabulary covers authentication, pagination, defaults, response transformation, type definitions, error mapping, examples, hints, coverage metadata, setup instructions, and composite tools.

Two structural decisions are worth noting. Tools are a YAML *map*, not an array, with the tool name as key; this enables direct reference and makes definitions more compact. Underscore-prefixed keys (`_defaults:`, `_common:`) are ignored by the runtime and exist solely as YAML anchor targets, allowing intra-file reuse without polluting the schema.

## 3.3 Authentication

DADL v0.1 supports five schemes that, in our experience, cover the vast majority of REST APIs:

- **Bearer token** — token injected as `Authorization: Bearer <token>`; header name and prefix configurable.
- **Basic** — username and password credentials combined and Base64-encoded by the runtime.
- **OAuth 2.0 client credentials** — token URL, client id/secret, scopes, token caching, and proactive refresh before expiry.
- **Session-based** — declarative login call with token extraction (JSONPath) and re-login on 401.
- **API key** — injected as a header or query parameter, with configurable name.

All credential references are logical names. The runtime is responsible for resolving them against whichever credential store is configured.

## 3.4 Pagination

Four strategies are supported, drawing on the vocabulary established by the Airbyte Low-Code CDK [11]:

| Strategy | Mechanism | Common in |
|---|---|---|
| `cursor` | Opaque cursor token in response | Stripe, Slack |
| `offset` | Numeric offset incremented by page size | Many generic REST APIs |
| `page` | Page number incremented per request | GitLab, Hetzner Cloud |
| `link_header` | RFC 8288 `Link: <…>; rel="next"` header | GitHub |

A pagination block specifies request parameter names, response extraction paths, a `max_pages` safety limit, and a `behavior` field with two values: `auto` (the runtime fetches all pages transparently) or `expose` (the LLM controls pagination explicitly via a cursor parameter). Pagination defaults are inherited and may be overridden per tool, including disabling it with `pagination: none`.

## 3.5 Response Transformation

API responses often contain far more data than is useful for the agent's context. DADL provides a four-step pipeline:

1. `result_path` — JSONPath expression extracting the relevant payload portion (e.g. `$.data` to unwrap an envelope).
2. `transform` — a jq filter applied after extraction, used for field projection, renaming, or reshaping.
3. `max_items` — array truncation.
4. `allow_jq_override` — when set, the LLM can pass an ad-hoc jq filter at call time for exploratory cases.

These transforms reduce token consumption substantially for large responses and normalize heterogeneous APIs into consistent shapes. The specification recommends defining a transform for any endpoint returning more than 5 KB per item.

## 3.6 Access Classification

Every tool may declare an `access` field with one of four well-known values:

| Level | Semantics |
|---|---|
| `read` | Retrieves data; no side effects |
| `write` | Creates or updates resources |
| `admin` | Privileged operations |
| `dangerous` | Destructive or irreversible operations |

Custom values (`billing`, `pii`, `ops`, ...) are passed through unchanged so that organizations can attach domain-specific policies. When `access` is omitted, no default is inferred, since inferring access from HTTP method alone is unsound: `POST /search` is a read operation, and `GET /admin/reset-cache` is dangerous despite being a GET.

The classification is a *contract* between the DADL file and the runtime's authorization system, not an enforcement mechanism in itself. DADL does not prescribe enforcement semantics; that is a property of the runtime and its policy engine. In particular, the format does not specify what happens when (a) a composite tool labelled `read` calls a primitive labelled `dangerous`, (b) a parameter binding promotes an apparently safe call into a privileged one, or (c) the file's author misclassifies a tool. These are policy-engine decisions and depend on the deployment. A conservative runtime should treat access labels as the *minimum* privilege required and apply the strictest label encountered along any composition path; a permissive runtime may evaluate the labels independently. Section 5.3 discusses the deployment model in the reference execution layer.

The format's responsibility ends at expressing the contract honestly: a tool that destroys data should be labelled `dangerous`, regardless of whether the policy is currently configured to enforce it.

### 3.7 Composite Tools

Some operations require combining data from multiple endpoints — joins, conditional logic, multi-step workflows — and cannot be expressed in a single HTTP request plus jq transformation. For these cases DADL provides **composite tools**: server-side JavaScript functions that invoke primitive tools through an `api.<tool_name>(params)` interface.

```
composites:
  get_named_status:
    description: "Get device status with human-readable names"
    params:
      only_on:
        type: boolean
        default: false
    timeout: 30s
    code: |
      const devices = await api.list_devices();
      const nameMap = Object.fromEntries(devices.map(d => [d.id, d.name]));
      const status = await api.get_all_device_status();
      const result = status.map(d => ({ ...d, name: nameMap[d.id] || d.id }));
      return params.only_on ? result.filter(d => d.relay_on) : result;
```

**Listing 2.** A composite tool joining device names from one endpoint with status data from another.

A second example, drawn from the public registry's `hackernews.dadl`, illustrates a different pattern:

```
composites:
  get_story_with_comments:
    description: "Get a story and its top-level comments resolved to full items"
    params:
      id: { type: integer, required: true }
      comment_limit: { type: integer, default: 20 }
    timeout: 30s
    code: |
      const story = await api.get_item({ id: params.id });
      if (!story || !story.kids?.length) return { ...story, comments: [] };
      const kidIds = story.kids.slice(0, params.comment_limit);
      const comments = await Promise.all(
        kidIds.map(id => api.get_item({ id }))
      );
      return {
        ...story,
        comments: comments.filter(c => c && !c.deleted && !c.dead),
      };
```

**Listing 3.** A composite from `hackernews.dadl`: fetch a parent item, resolve its child IDs in parallel, and filter out deleted entries.

Across the public registry, composite tools cluster into four recurring patterns:

- **Join across endpoints** — combining fields from two related calls (Listing 2).
- **Lookup-then-resolve** — fetching a list of IDs and resolving each to its full record, often in parallel

(Listing 3).

- **Conditional dispatch** — choosing between alternative endpoints based on a parameter, e.g. relevance vs. recency sort.
- **Projection for context efficiency** — stripping API-internal metadata before returning to the LLM. The `search_distilled` composite in `algolia-hn-search.dadl` reduces approximately 2 KB per result to roughly 200 bytes by removing fields such as `_highlightResult`, `_rankingInfo`, and `_tags` that the LLM never uses.

Each pattern captures repeated work the team has already done by hand and makes it available to every agent thereafter, without altering the underlying primitive tools.

Composite code runs in a restricted sandbox. External I/O (`fetch`, `require`, `import`, filesystem access) is forbidden, as are dynamic-evaluation primitives (`eval`, `Function`). Execution is hard-killed after a configurable timeout (default 30 s, maximum 120 s), and a maximum number of underlying API calls per execution prevents runaway loops. Every primitive call inside a composite is logged with the composite name as parent context. Files containing composites are flagged `contains_code: true` so that automated review pipelines can apply additional static analysis.

These restrictions are engineering measures, not a formal sandbox model. We do not claim a security proof: the sandbox is implemented with the JavaScript engine's standard isolation primitives plus a static-analysis pass over the submitted code, and its robustness against adversarial composites is an open question (Section 9). The right interpretation is that the declarative path through DADL retains Property 1 (Declarative Closure) by construction, and composites are a deliberate, scoped escape hatch whose security properties depend on the runtime's sandbox implementation rather than on the format itself.

## 3.8 Hints

A frequently underestimated cost of integrating a real API is *institutional knowledge*: the small facts that are not in the OpenAPI spec but determine whether a call succeeds. DADL captures these in a per-tool `hints` block:

```yaml
hints:
  list_project_tasks:
    position_type: float64
    requires: "call list_views first to get view_id"
    kanban_note: >
      Kanban views return buckets with nested tasks, not a flat list.
```

Hints are injected into the tool description at load time. Because they live next to the tool definition rather than in a separate wiki, they do not drift from the API as it evolves and are version-controlled like the rest of the file.

## 3.9 Coverage Metadata

Every backend may declare a `coverage` block with the number of tools defined, the estimated total of endpoints in the upstream API, a percentage, free-text fields for *focus* and *missing* areas, and a `last_reviewed` date. This is documentation for human reviewers and a planning artefact for incremental contribution: gaps in coverage are explicit and addressable.

The GitHub DADL in the public registry, for example, declares 205 tools out of an estimated 900 GitHub REST endpoints (≈23 %), with explicit lists of covered and uncovered areas. This lets a contributor extending the file see immediately what is already done and what is not.

## 3.10 Error Handling and Rate Limiting

DADL declares error format (JSONPath to message and code), terminal status codes, retryable status codes, an exponential-backoff retry policy, and a proactive rate-limit configuration. When the rate-limit block is present, the runtime monitors `X-RateLimit-Remaining` and `Retry-After` headers and pauses requests before exhausting the quota rather than waiting for `429` responses. This is particularly important when many users share the same API credential, which is the typical case in an enterprise tool library.

## 3.11 Property: Declarative Closure

The structural choices above support a property of the format itself, which we state explicitly:

> **Property 1 (Declarative Closure).** For any DADL file F, the set of HTTP endpoints any tool defined in F can address is finite and determinable by static inspection of F alone. The dynamic *which-and-how-many-times* of those endpoint accesses depends on inputs and (for composite tools) on data; the *which-endpoints-are-reachable* does not.

Three points are worth making about what this property does and does not claim.

**It bounds, it does not summarize.** Declarative Closure says the HTTP surface is finite and statically determinable. It does not say a human can comprehend a 2,700-line GitHub DADL at a glance — large files still require careful reading. The value is that the reading is *terminating*: a reviewer who has finished the file has finished the review of HTTP behavior. There is no hidden code path that can construct a request the file does not name.

**Composite tools are an explicit escape hatch.** A composite's JavaScript can call primitive tools in arbitrary order, with arbitrary parameter values derived from data. Declarative Closure still holds — the composite can only call tools defined in F, and each such tool has its own statically-determinable HTTP surface — but the *composition* of calls is dynamic. Files using composites are flagged `contains_code: true`; reviewers should treat them as code, not configuration.

**Hand-written MCP servers cannot offer this property in general.** Arbitrary code may construct request URLs at runtime, conditional on inputs, environment, or remote responses. Declarative Closure is the technical underpinning of what we informally describe as *auditable HTTP surface* (Section 3.1). It is a property of the format, not a property of any particular DADL file's quality.

---

# 4. Code Mode

## 4.1 The Problem

The standard MCP convention registers each tool individually. For an organization with N APIs averaging M endpoints each, the LLM's context cost for tool advertisement is O(N·M). With realistic N (10–30) and M (10–200), the cost is large enough to dominate the available context window and incompatible with the goal of broad organizational coverage.

## 4.2 The Mechanism

DADL backends are exposed to the LLM through exactly two MCP tools:

- `search` searches the available catalog by keyword and returns a small set of matching tool signatures with descriptions.
- `execute` runs a JavaScript snippet against an auto-generated TypeScript interface, where each DADL tool becomes a method (`api.list_repos({...})`).

The LLM thus *discovers* tools when needed and *uses* them as code rather than as protocol calls. The runtime parses the DADL files, generates TypeScript signatures from the parameter and response definitions, and exposes them only inside the `execute` sandbox. A typical DADL backend's full TypeScript definition is generated automatically and is never sent into the conversation.

The cost of having the catalog available is therefore approximately O(1) in catalog size: only the descriptions of `search` and `execute` themselves and a brief introduction are visible. With the public registry at ~1,500 tools, this is the difference between a fully usable agent and one that can no longer fit a user's question.

## 4.3 What the LLM Writes

Code Mode shifts the unit of agent action from a single tool call to a small JavaScript snippet. A typical snippet looks like:

```
const issues = await api.list_repository_issues({
  owner: "DunkelCloud",
  repo: "ToolMesh",
  state: "open",
  labels: "bug",
});
return issues.filter(i => !i.assignee).map(i => ({
  number: i.number, title: i.title, url: i.html_url,
}));
```

Side effects, intermediate filtering, and joins live in the snippet. The runtime enforces the same sandbox restrictions as for composite tools: no external I/O, no dynamic evaluation, hard timeout, capped number of underlying API calls. A single `execute` invocation therefore behaves like a controlled, audited mini-program rather than an open shell.

## 4.4 Trade-offs

Code Mode trades the simplicity of one-call-one-tool for two effects: bounded tool-advertisement context cost and a richer per-call vocabulary. The cost of reasoning about a JavaScript snippet is borne by the model itself; current frontier models handle this routinely, and the ergonomic benefit of being able to filter, reshape, and combine in a single round trip outweighs the cognitive cost in our experience. The runtime requires a sandboxed JavaScript engine, which is an implementation cost paid once.

Three further trade-offs are less obvious and deserve explicit mention.

**Tool-call inspection moves from the protocol to the runtime.** In conventional MCP, every tool invocation is a typed call with a fixed schema; clients, gateways, and audit systems can inspect, validate, and approve each call without executing it. In Code Mode, the unit of action is a JavaScript snippet that the model produces at request time. Pre-execution inspection is still possible — the runtime sees the snippet before it runs — but it is now structural code analysis rather than schema matching, and downstream consumers of MCP traces (UIs, approval prompts, observability tools) must be adapted to log primitive `api.*` calls instead of top-level tool calls.

**Prompt injection's surface shifts.** In conventional MCP, an injected prompt that smuggles a tool name attempts to invoke a known tool; the authorization check still applies. In Code Mode, an injected prompt may smuggle JavaScript that constructs unexpected primitive calls. The authorization check still applies at the primitive level (each `api.*` call passes through the same access-classification gate), but the surface for *which sequence* gets executed is broader. Designs that rely on tool-name allow-listing must move to access-level allow-listing.

**Native MCP tool granularity is lost.** Some MCP clients, UIs, and approval flows assume each operation is its own tool with its own metadata. Code Mode replaces N tools with two (`search`, `execute`); per-tool permissions, per-tool documentation pages, and per-tool client affordances are no longer addressable from outside the runtime. A runtime can re-expose individual DADL tools as MCP tools when needed (the format does not preclude it), but doing so reintroduces the O(N) advertisement cost the design was meant to avoid.

These are real costs, paid in exchange for the constant-cost advertisement of Property 2 (Section 4.5). Whether they are worthwhile depends on the deployment: an agent meant to run autonomously over a large internal API surface benefits more than an agent meant to expose two or three carefully-curated tools to an end user.

## 4.5 Property: Constant-Cost Tool Surface

The defining property of Code Mode can be stated formally:

> **Property 2 (Constant-Cost Tool Surface).** Under Code Mode, the LLM context cost for tool advertisement is O(1) in the size N of the DADL backend catalog. The TypeScript interface for any tool is materialized only inside the `execute` sandbox upon retrieval via `search`, not in the conversation context.

The constant is small. On the 1,833-tool public registry (Section 6.2), the empirical advertisement cost is approximately 1,000 tokens, independent of how many backends are in scope. The same catalog under conventional one-tool-one-schema MCP advertisement consumes approximately 142,000 tokens — a 142x reduction. Constant-Cost Tool Surface is the property that turns "the agent has access to many APIs" from an aspiration into an engineering reality.

---

# 5. The Enterprise Tool Library Pattern

DADL's individual features — declarative description, central runtime, access classification, hints, coverage — combine into a pattern that we have found to be the most consequential outcome of the design: an organization-wide, versioned, governable library of API integrations. This section describes that pattern.

## 5.1 Definition vs. Server

The minimal artefact for adding an API to the library is a YAML file. There is no Dockerfile, no dependency manifest, no CI pipeline, and no service entry to register. A new integration is a single pull request that adds one file to a directory.

This collapses the cost of contribution dramatically. In an organization where a developer must justify an additional service in their cluster every time they want to integrate a new API, integrations are scarce and concentrated in a small operations team. With DADL, the developer who needs the integration writes the file, opens a pull request, and the integration becomes available to everyone after review and merge. The maintenance unit is no longer "a service" but "a configuration file", with all the version-control, code-review, and rollback affordances that follow.

## 5.2 Centralized Authentication

The most concrete operational benefit of consolidating all REST integrations into a single runtime is that *credentials* are managed in one place. In the conventional model, each MCP server stores or fetches its own credential. A GitHub PAT exists in one server, a Stripe secret key in another, and SSO tokens fan out across as many servers as require them. Rotation, revocation, scope auditing, and incident response must be repeated for every server.

In an enterprise tool library, every DADL file references credentials by logical name. The runtime owns the credential store and resolves references at request time. Rotating a credential is a one-place operation. Revoking access to a service is a one-place operation. Producing an audit log of which credentials were used by which user against which tool is a one-place operation. This is not a feature of DADL itself but of consolidating *all* integrations behind one runtime; DADL's contribution is making such consolidation tractable, since each new integration is a YAML file rather than a server.

## 5.3 Centralized Authorization

The four well-known access levels (Section 3.6), combined with arbitrary custom values, give the authorization layer a uniform vocabulary across the entire tool catalog. A policy that says "members of the on-call team may invoke tools with `access: dangerous` only in incidents tagged P1" can be expressed once and applied to every DADL backend without modification of the DADL files themselves. By contrast, when authorization logic lives in individual MCP servers, each server is its own policy engine, with its own bugs and its own update cadence.

This matters at the scale of a real library. With ten tools, an organization can review and approve each one individually; with fifteen hundred, it cannot. The four-level vocabulary lets the entire catalog be governed by a small number of policies — "every employee may invoke `read` tools", "engineers may also invoke `write`", "on-call may invoke `dangerous` during active incidents" — that apply automatically to each new DADL file the moment it joins the library, with no policy change required. Custom levels (`billing`, `pii`, `ops`) extend the same mechanism to domain-specific concerns: a `pii` policy is written once and applies retroactively to every existing tool that carries the tag and prospectively to every new tool that declares it. The classification therefore turns "who may do what" from an O(tools) decision into an O(roles) decision.

The reference execution layer enforces the contract at three points: when an agent invokes `execute`, the runtime parses the JavaScript and identifies the primitive `api.*` calls; each primitive call is checked against the caller's role and the tool's `access` label *before* the HTTP request is issued; and the result, including any access denial, is recorded in an audit log keyed by caller, tool, and access level. Access checks therefore happen per primitive call, not per `execute` invocation: a `read`-labelled composite that internally calls a `dangerous` primitive will see the dangerous call denied (or approved) on its own merits, regardless of the composite's outer label. A conservative runtime may additionally refuse to execute composites whose static analysis reveals a privilege escalation path that the caller would not be authorized to take directly; the format permits this but does not mandate it.

The reference execution layer uses OpenFGA [12] as its authorization engine; the choice of engine is orthogonal to DADL and reflects a preference for relationship-based access control [13]. What matters for the library pattern is that the decision is made *outside* the DADL file: the file declares the access level, the policy decides who may invoke it.

## 5.4 Coverage as a First-Class Concept

A library's value is bounded by what it covers. The `coverage` block (Section 3.9) elevates this from an implicit property to an explicit, machine-readable field. In our public registry, the GitHub DADL declares 205 of 900 endpoints (23 %), the NetBox DADL declares 222 of 250 (≈89 %), and the DeepL DADL declares all 21 of its endpoints (100 %). These numbers are visible in the registry and in the tool's documentation, allowing an organization to identify which APIs are well covered, which are sparse, and where contribution effort is most useful.

Coverage tracking also addresses what we informally call the **shadow API problem**: APIs that internal applications already call but that are not part of the agent-accessible library. Making coverage explicit gives the library curator a backlog of APIs to bring under common authentication and authorization control.

## 5.5 The Library Composition Effect

A library's value is not the sum of its parts. We name the **Library Composition Effect** the emergent capability of an LLM agent, given access to a declaratively-described tool library spanning multiple sources behind a single authentication and authorization boundary, to construct cross-source queries that no individual tool could answer. The effect is not a feature of any one DADL file; it appears only at the *library* level, when many such files share a runtime, a credential perimeter, and a uniform Code Mode interface.

A typical infrastructure diagnosis spans multiple sources: an alert from Alertmanager identifies an affected service; topology data from NetBox locates the host; recent logs from Graylog reveal the failure mode; the runbook entry in a wiki proposes a recovery; the hypervisor state in Xen Orchestra confirms what the alert reflects; the cloud-side instance metadata from a provider like Linode or Hetzner Cloud completes the picture. In Code Mode (Section 4), this composition is expressed as a short program the LLM constructs in response to the question, against tools from any DADL backends in scope, with one set of credentials and one audit trail.

This is the practical end of the "broad coverage" argument: not "the agent has access to more APIs" but "the agent can answer questions that span sources". Without a unified library, the same investigation requires either separate sessions per data source — with no shared context — or a per-task integration that re-implements the same join each time. Both fragment the agent's view. The library pattern makes such composition cheap because the marginal cost of including one more source is a single DADL file.

Critically, the composing program is not written by the library author. It is constructed by the LLM at request time, against whichever subset of the library is in scope, in response to a question that may never have been anticipated. Composite tools (Section 3.7) are the inverse — fixed compositions the library author has decided are worth encoding once. The two mechanisms are complementary: composites capture the workflows the team already knows it needs; Code Mode handles everything else.

## 5.6 Encoding Institutional Knowledge

The hints mechanism (Section 3.7) lets an organization's tribal knowledge live next to the tool definition. *"Set* `position_type` *to float64, not integer." "Call* `list_views` *first to get a* `view_id`*." "Kanban views return buckets, not flat lists."* These facts are usually communicated through onboarding conversations, Slack threads, and stale wiki pages, and they are the most common source of repeated agent failure when an integration is otherwise sound.

By living in the DADL file, hints are version-controlled, reviewable, attributable, and visible to every agent that uses the tool. New hints accumulate as the team's experience with the API grows. Composite tools (Section 3.7) play the same role for *workflows*: a sequence of calls that the team has learned is the correct way to perform a common operation can be encoded once and made available to every agent thereafter.

## 5.7 Operational Properties

Treating the library as configuration rather than code yields several operational properties.

**Hot reload.** A DADL file can be updated and re-read without restarting the runtime. The unit of deployment is a YAML diff, not a container image.

**GitOps friendly.** Because DADL files are plain YAML and contain no secrets, the canonical store is a git repository. Branching, code review, signed commits, and CI-driven validation apply directly. Promotion from staging to production is a `git merge`.

**Auditable.** Each invocation produces a structured log entry referencing the DADL file, the tool, the access level, the calling user, and the resolved credential identifier (not value). Combined with central auth/authz, this gives a single timeline of who did what across the entire API surface.

**Reviewable by non-developers.** Because the format is declarative and free of secrets, security reviewers, compliance officers, and domain experts can read DADL files without programming-language fluency. This widens the pool of competent reviewers, which matters in regulated environments.

## 5.8 Library Authoring Workflow

Producing a DADL file from API documentation is, in our experience, a 5-to-60-minute task with LLM assistance: provide the API documentation (or OpenAPI spec) and the DADL specification, and ask for a draft. The result typically requires one or two correction cycles for non-obvious authentication or pagination patterns. We have used this workflow to produce most of the public registry's content.

Once a DADL file exists, the marginal cost of extending it is lower still. Adding ten endpoints to an existing 100-endpoint file is a matter of copying tool blocks and adjusting paths, parameters, and access levels. A coverage block that says "23 % covered" is also a checklist for further work.

**Agentic Toolsmithing.** A pattern we have found unusually productive is to ask the agent that just used a DADL whether the integration helped or hindered the task: which tools were missing, which response shapes were too verbose, which hints were absent or wrong, which composite would have collapsed three calls into one. The agent's answer — often more incisive than a human reviewer's, because it has just felt the friction — is folded back into the DADL file. Many of the hints, response transforms, and composite tools in the public registry originated this way. Because the agent is at once consumer and critic of the library, the library improves as a function of its own use.

We borrow the term *toolsmithing* from Brooks [15], who described the computer scientist who builds and refines the tools they themselves rely on. The agentic variant extends this tradition to the case where the LLM agent participates directly in the refinement loop, not as the smith's apprentice but as the user whose feedback drives the next iteration.

This loop is only practical because the artifact is a single declarative file. Iterating on a hand-written MCP server in the same way would require a code change, a rebuild, and a redeploy per iteration; iterating on a DADL file is a YAML edit and a hot-reload. The cost of capturing the agent's feedback approaches the cost of the feedback itself.

---

# 6. Practical Properties

This section gives concrete numerical evidence for DADL's claims drawn from the public registry at https://dadl.ai. We do not present a head-to-head benchmark against MCP servers, since the relevant comparison is operational rather than algorithmic and depends heavily on which MCP server implementation is chosen as the baseline.

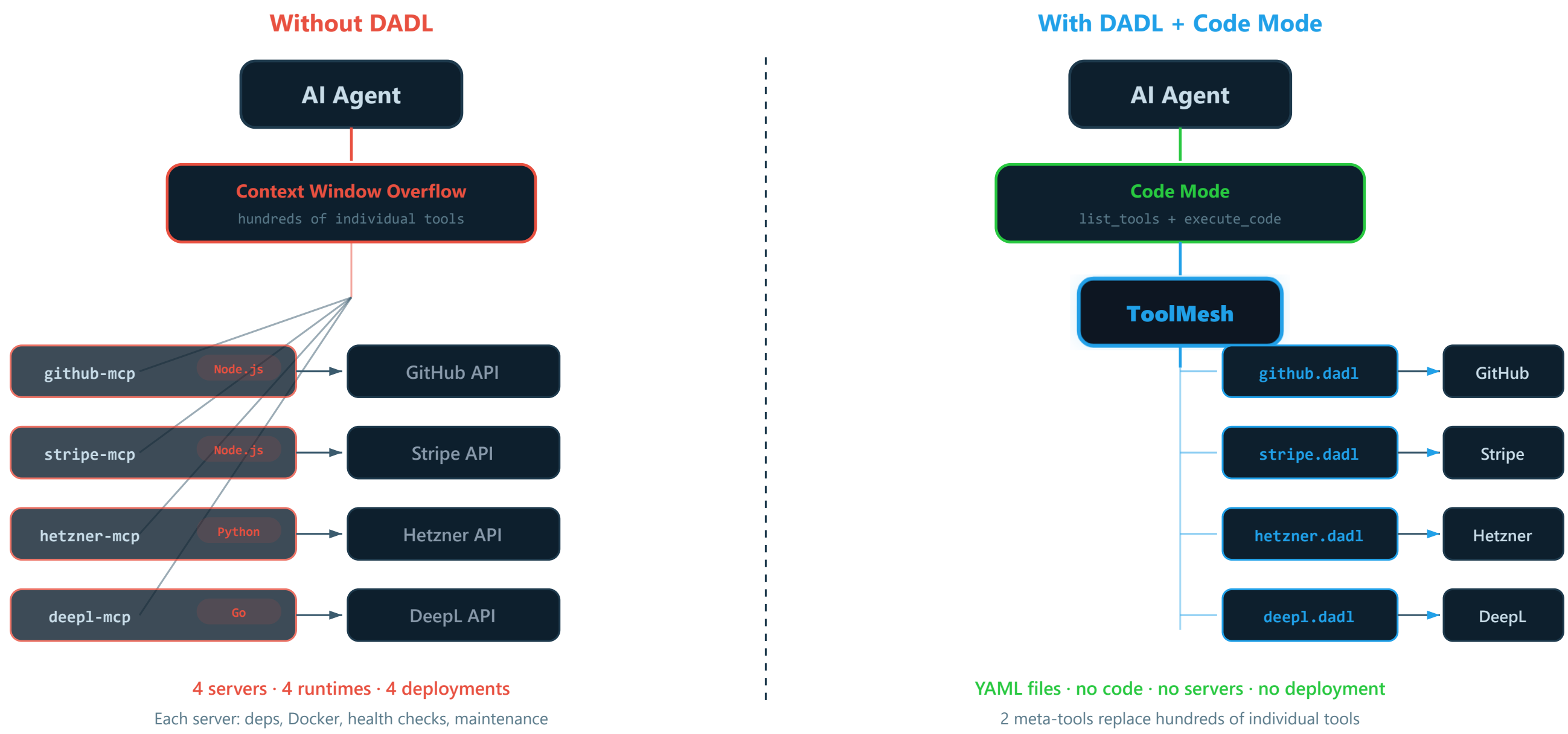

**Figure 1.** Side-by-side architectural comparison of the conventional one-MCP-server-per-API model (left) with the DADL + Code Mode model (right). The left side shows N independent server processes, each with its own runtime and deployment, exposing NxM individual tool schemas that overflow the LLM context window. The right side shows a single shared runtime that interprets per-API DADL files; the agent reaches the entire catalog through two meta-tools (`list_tools`, `execute_code`). The SVG is included in the paper repository at `figures/dadl-flow.svg`.

## 6.1 Registry Snapshot

As of April 2026, the public registry contains 20 DADL files covering services that include GitHub, GitLab, Stripe, Cloudflare, NetBox, Mastodon, Hetzner Cloud, Hashicorp Nomad, Linode, Tailscale, Vikunja, DeepL, Hacker News, and others. Total size is approximately 27,000 lines of YAML, exposing 1,833 tools. The median DADL covers approximately 80 % of its upstream API; median lines-per-tool is 13. Selected per-API figures:

| API | Tools | YAML lines | Auth | Pagination |
|---|---:|---:|---|---|
| NetBox | 222 | 3,002 | Bearer | offset |
| GitHub | 207 | 2,694 | Bearer | link_header |
| Cloudflare | 195 | 2,587 | Bearer | cursor |
| Hetzner Cloud | 139 | 1,584 | Bearer | page |
| Mastodon | 126 | 1,577 | Bearer | link_header |
| Stripe | 107 | 1,539 | Bearer | cursor |

| API | Tools | YAML lines | Auth | Pagination |
|---|---:|---:|---|---|
| GitLab | 79 | 1,181 | Bearer | page |
| Xen Orchestra | (varied) | 1,244 | Session | offset |
| DeepL | 21 | 478 | API key | none |
| Vikunja | 20 | 338 | Bearer | offset |

**Table 1.** Selected DADL definitions in the public registry.

The line-to-tool ratio averages around 12–15 lines of YAML per tool, including the per-tool description, parameter list, access level, and any tool-specific overrides for pagination, response shaping, or error mapping. The overhead at the file level (auth, defaults, types, examples, hints, coverage, setup) is amortized across all tools in the file.

## 6.2 Context Window Behavior

A DADL backend of any size, when exposed via Code Mode, contributes the same fixed overhead to the conversation context: the descriptions of `search` and `execute` and a short orientation prompt, totaling roughly 1,000 tokens. The TypeScript interfaces generated from the DADL file are not in the conversation context; they are made visible to the model only inside the `execute` sandbox. An agent with access to the entire public registry therefore sees no more tool advertisement than an agent with access to a single backend.

We measure this directly. On the public registry's 1,833 tools across 20 backends, the full TypeScript interface payload is approximately 634 KB, or about 214 bytes per tool when retrieved through `search`. Encoded as a flat MCP tool-registration footprint (each tool's name, description, and JSON Schema for parameters in the conversation context), the equivalent advertisement consumes approximately 142,000 tokens. Code Mode replaces this with the fixed ~1,000-token surface — a **142x reduction for the full catalog**, and a **5.9x reduction at the median backend** (~92 tools). At median, the full Code Mode advertisement is smaller than a single conventional MCP server's tool list.

This is the property that makes a true library practical at all. Without it, the question "how many APIs can my agent know about" has an answer measured in single digits.

## 6.3 LLM-Assisted Authoring

We routinely produce new DADL files by giving a current LLM the API's OpenAPI spec or HTML documentation, the DADL v0.1 specification, and one or two existing DADL files as examples. For well-documented APIs, a usable first draft is returned in a single interaction. For APIs with idiosyncratic authentication or pagination, one to three correction cycles are typical. Time-to-first-working-tool for a previously unsupported API is consistently under one hour.

## 6.4 Comparison

For completeness, we summarize how DADL relates to the two adjacent approaches discussed in Section 2.

| | DADL | Hand-written MCP server | AutoMCP |
|---|---|---|---|
| Artefact | YAML | Code (TS/Python) | Generated code |
| Process per API | None (shared) | One | One |
| Adds a new API | One YAML file | Project + deploy | Generation + deploy |
| Context cost | O(1) via Code Mode | O(N) tools | O(N) tools |
| Customization | Edit YAML | Edit code | Edit generated code |
| Auth/authz model | Central (runtime) | Per-server | Per-server |
| LLM-generatable | Yes (from docs) | Partial | Partial (needs OpenAPI) |

**Table 2.** Comparison with adjacent approaches.

DADL is best understood as complementary to native MCP servers rather than as a replacement. Integrations that require WebSocket connections, custom request signing, or arbitrary computation are not DADL's target. For the dominant REST request-response pattern that constitutes the bulk of current integrations [4], the declarative model removes a class of operational work.

## 6.5 The Registry as a Research Artifact

The dadl.ai registry is, to our knowledge, the largest uniform machine-readable corpus of REST-API descriptions tailored for LLM tool use. We release it under CC BY-SA 4.0 as a research artifact for downstream work in:

- **REST-to-DADL translation benchmarks.** Given an OpenAPI specification or HTML documentation, produce a syntactically valid DADL file. The 20 services in the registry serve as ground-truth references with declared coverage percentages.
- **Cross-API tool-use research.** A uniformly-described catalog of 1,833 tools across 20 services supports controlled studies of LLM tool selection, decomposition, and the Library Composition Effect (Section 5.5) at non-trivial scale.
- **Empirical characterization of API integration patterns.** Quantitative analysis of authentication-scheme distribution, pagination prevalence, response-shape complexity, hint usage, and composite-tool patterns across services.
- **LLM context-cost measurement.** The 142x/5.9x comparison of Section 6.2 is reproducible from a registry checkout; finer-grained measurements per backend, per auth type, per pagination strategy are directly available.
- **Adversarial sandbox studies.** Composite tools in the registry are real-world examples; adversarial variants can be added to the corpus to evaluate sandbox robustness (Open Question 5, Section 9).

The registry is versioned (git, with semantic versioning per DADL file), reproducible (each file declares its date of last review and a coverage percentage), and open to community contributions through the project repository.

# 7. Discussion

## 7.1 What DADL Is Not

DADL is deliberately scoped. It is not a workflow engine: there are no triggers, schedulers, or persistent state machines in a DADL file. It is not an authorization engine: it declares access *levels*, but the policy is owned by an authorization system. It is not a credential vault: it holds *references*, not values. These exclusions are what make DADL files small enough to be readable and reviewable by non-developers.

The composite-tool feature is the single deliberate carve-out for executable content, and it is bounded by an explicit sandbox and a `contains_code: true` flag. We considered making composites a separate format and decided against it on the grounds that the cost of an additional file format would exceed the cost of carefully scoping the existing one.

## 7.2 Limitations

Three classes of API are not well served by DADL.

**APIs requiring custom request signing.** AWS Signature Version 4, certain GraphQL signatures, and similar schemes require arbitrary computation per request. DADL provides no extension point for this and intentionally does not aim to.

**APIs centered on persistent connections.** WebSocket-based streaming and bidirectional protocols do not fit the request-response model that DADL describes.

**APIs whose practical use requires substantial application-side logic.** When an "API integration" is in fact 70 % business logic and 30 % HTTP, a code-based MCP server is the right tool.

For these cases, native MCP servers remain appropriate. A runtime that supports both DADL and native MCP backends behind the same authorization, auditing, and credential isolation is straightforward to construct, and we view this hybrid as the realistic deployment shape.

## 7.3 Risks of the Declarative Model

A purely declarative format can fail to capture an API faithfully. Two failure modes are worth naming.

**Under-expressive vocabulary.** If an API uses a pagination scheme not covered by the four DADL strategies, a DADL file cannot describe it. We have not encountered such a scheme in the public registry, but the possibility exists. The mitigation is to extend the vocabulary in subsequent revisions or fall back to a native MCP server for that specific API.

**Drift between the spec and the upstream API.** A DADL file describes the API at the time of authoring. APIs evolve. The `version`, `last_reviewed`, and `coverage` fields exist precisely so that drift is visible; they do not prevent it. This is no worse than the equivalent problem for hand-written MCP servers and code-generated wrappers, but it is also no better.

## 7.4 Audit and Compliance Implications

For organizations subject to regulatory or contractual auditing requirements, the declarative shape of a DADL library has measurable consequences. A complete inventory of "every external API call our agents may make" is a directory listing. A complete inventory of "every privileged operation" is a query for `access: admin` or `access: dangerous` across that directory. A complete history of "what changed last quarter" is a `git log`.

None of these properties hold for an MCP-server fleet without considerable additional infrastructure.

---

## 8. Related Work

Hou et al. [3] provide the first systematic landscape study of the MCP ecosystem and characterize a four-phase server lifecycle (creation, deployment, operation, maintenance) with associated security threats. Mastouri et al. [4] establish the empirical case for the wrapper pattern and demonstrate automated server generation. Hasan et al. [5] document maintainability and security weaknesses across the broader server population, and Radosevich & Halloran [6] demonstrate exploitable vulnerabilities in deployed MCP servers including credential leakage and command injection. Li et al. [8] generate tool definitions from unstructured documentation. MCP Bridge [9] aggregates servers behind a proxy. Lumer et al. [7] address the context-window cost of large tool catalogs from a complementary angle, treating tool selection as a short-term-memory management problem at the LLM client.

Beyond the MCP ecosystem, DADL is best understood against several adjacent traditions of declarative API description (Section 2.4). OpenAPI [10] describes APIs for human and machine consumers but lacks the LLM-oriented affordances DADL adds. Terraform's provider model is the closest operational analogue: declarative resources, runtime-resolved credentials, a shared runtime, and a community provider registry; DADL targets request-driven semantics rather than desired-state reconciliation. The Airbyte Low-Code CDK [11] supplied DADL's pagination vocabulary. API gateway configurations (Kong, Apigee, Tyk, AWS API Gateway) declare routing, authentication, and transformation in YAML or JSON for client-side exposure; DADL inverts the direction, declaring how an LLM-side agent calls an upstream API. Postman Collections capture executable HTTP requests for human-driven exploration but rely on imperative scripting and so do not satisfy Property 1 (Section 3.11). AsyncAPI describes event-driven APIs and is orthogonal to DADL's REST request-response scope.

The relationship-based access control model [13] underpinning OpenFGA [12] influenced the choice to keep authorization decisions outside the DADL file and inside an engine that can express organization-wide relationships rather than per-tool ACLs.

Brooks [15] introduced the figure of the *toolsmith* — the computer scientist who builds and refines the tools they themselves rely on. Section 5.8 extends this lineage to the agentic case, where the agent that uses a DADL file also authors its refinements.

---

# 9. Open Questions

The DADL specification raises questions it does not resolve. Empirical work on scientific influence finds that the most consequential papers in a field tend to be those that *open* questions in their concept network rather than merely answering existing ones [14]; in that spirit, we close this paper by stating, explicitly, the questions DADL raises.

**The Library Composition Effect (Section 5.5).** We have named the phenomenon and given a qualitative example. Its mechanism, scaling properties, and limits are uncharacterized. How does an agent's accuracy in cross-source decomposition scale with library size and heterogeneity? What tool-description granularity maximizes composability? Is there a threshold beyond which a library's value grows super-linearly with its size, or is the relationship monotonic but bounded? Is composability *transitive* — i.e., do agents reliably chain three or four sources, or does accuracy collapse beyond two?

**Coverage as a Cross-Organizational Metric (Section 3.9, 5.4).** Aggregated across an organization, the gap between APIs in operational use and APIs reachable through the agent library is a measurable property — what we have informally termed the *shadow API problem*. We are not aware of prior work proposing or measuring this metric across organizations. What is the typical coverage gap in real deployments? How does it correlate with operational incidents, audit findings, or developer toil?

**The Auditability–Expressiveness Frontier (Section 3.7).** DADL excludes arbitrary computation from its base vocabulary, with composite tools as a small, explicitly-marked carve-out. We have no theory of where the boundary should lie. What fraction of REST APIs in current MCP-server use is fully expressible in DADL without composites? Are the four composite patterns (join, lookup-then-resolve, conditional dispatch, projection) sufficient for a definable fraction of cross-endpoint workflows, or are additional patterns needed? Can the boundary be characterized formally, e.g. in terms of a complexity class of jq expressions?

**Hint Efficacy (Section 3.8).** The hints field embeds institutional knowledge alongside tool definitions. Whether — and how much — this improves LLM tool-use reliability is an open empirical question. A controlled study comparing tool-use accuracy with and without hints, on a defined task suite drawn from the public registry, would directly answer it.

**Sandbox Granularity for Composite Tools (Section 3.7).** The sandbox blocks external I/O and dynamic evaluation, with hard timeout and call-count limits. Whether this granularity is correct — too restrictive for legitimate workflows, too permissive against hostile DADLs — is open. A formal threat model for composite execution, together with adversarial evaluation against a corpus of malicious DADL submissions, would be a useful contribution.

These questions are deliberately not framed as our future work. The public registry at [https://dadl.ai](https://dadl.ai) provides a uniform corpus of REST-API descriptions in a single machine-readable format and is, to our knowledge, the largest such corpus currently available; it is offered as a starting point for empirical study of any of the questions above.

## 10. Conclusion

The MCP ecosystem has succeeded as a protocol and is now hitting the limits of its initial deployment shape. Servers proliferate, contexts saturate, credentials and authorization fragment across processes, and the cost of *adding* an API to an organization's reach is high enough that coverage stays narrow.

DADL replaces the per-API server with a per-API YAML file interpreted by a shared runtime. The change is not primarily technical — interpreted REST descriptions are an old idea — but organizational. When integrations are configuration rather than code, an organization can build a true tool library: versioned, auditable, governable, contributed to by anyone with API knowledge, served behind a single authentication and authorization boundary, and reachable by an agent through a Code Mode interface whose cost is independent of catalog size.

The DADL v0.1 specification is available under CC BY-SA 4.0 at [https://dadl.ai](https://dadl.ai). The public registry currently contains over 1,500 tool definitions across 20 services and is open to community contributions.

---

# Acknowledgments

Drafting assistance for this paper was provided by Claude (Anthropic). The author takes full responsibility for all technical content, architectural decisions, and conclusions.

**Cite as:** Dunkel, A. (2026). *DADL: A Declarative Description Language for Enterprise Tool Libraries in LLM Agent Systems*. Zenodo. https://doi.org/10.5281/zenodo.19931788